\documentclass{article}
\usepackage[]{graphicx}
\usepackage[]{epsfig}

\newcommand{\beq}{\begin{equation}}
\newcommand{\eeq}{\end{equation}}
\newcommand{\beqa}{\begin{eqnarray}}
\newcommand{\eeqa}{\end{eqnarray}}
\newcommand{\non}{\nonumber}

\begin{document}

\title{On Supersymmetry Breaking in the Computation of the Complexity}
\author{G. Parisi and T. Rizzo
\\
\small Center for Statistical Mechanics and Complexity, INFM Roma ``La Sapienza'' and 
\\
\small Dipartimento di Fisica, Universit\`a di Roma ``La Sapienza'', 
\\
\small Piazzale Aldo Moro 2, 00185 Roma, 
Italy}

\maketitle
\begin{abstract}
We study the consequences of supersymmetry breaking in the computation of the number of solutions of the Thouless-Anderson-Palmer (TAP) equations. We show that Kurchan argument that proves the vanishing of the prefactor of the Bray and Moore saddle point for the total number of solutions can be extended to solutions at any given free energy. We also provide a new simple argument for the vanishing of the prefactor and use it to prove that the isolated eigenvalue recently considered by Aspelmeier, Bray and Moore is exactly zero in the BM theory because of supersymmetry breaking. The behavior of the eigenvector of the isolated eigenvalue at the lower band edge is also considered.  
\end{abstract}
 
\section{Introduction}

The Complexity, defined as the logarithm of the number of metastable states divided by the size of the system $N$, is a key concept in disordered system theory. Recently, the classical Bray and Moore (BM) computation of the complexity in the Sherringhton-Kirkpatrick (SK) model \cite{BMan} has been reconsidered \cite{CGPM,noian,noiquen,ABM} starting from the observation that it violates a supersymmetry (SUSY) as noted in \cite{K91}. In this work we discuss some consequences of SUSY breaking.

Stable and metastable states of the system can be associated to solutions of the Thouless-Anderson-Palmer (TAP) equations \cite{TAP} but
in general not any TAP solution can be actually identified with a thermodynamical state, see {\it e.g.} discussion in \cite{Ple1,noian}, this point however is not important to the present discussion therefore for the sake if simplicity we will identify the complexity as the logarithm of the total number of TAP solutions.
Much more important is the fact that for technical reasons the actual object we consider is not the sum of all solutions of the TAP equations but rather a sum where each solution is weighted with the sign of determinant of the Hessian of the TAP free energy. This modification is very important and it is the very origin of the SUSY of the problem.
The first consequence of this modification is that the weighted sum over all solutions must be trivially equal to one, indeed the Morse Theorem states that the total number of solutions of the TAP equations with positive determinant minus the number of solutions with negative determinant is equal to one. 
Therefore the problem splits in two: from one hand we compute the sum without the modulus but on the other hand we want to extract from the very same computation the actual complexity, {\it i.e.} the sum {\it with} the modulus.
In other words we want to compute the total complexity but we want also to recover the trivial result that the sum without the modulus of all solutions be equal to $1$.

In the following we briefly outline the results of the paper.
In section \ref{sec:compesup} we will express the number of solutions of the TAP equations as an integral over the exponential of an action depending on bosonic and fermionic (Grassmanian) variables. Due to the mean-field nature of the model this integral can be evaluated for large $N$ by the saddle point method. Then the prefactor of the leading exponential contribution can be obtained as a power series in $1/N$. The problem is to solve the saddle point equations obtained in order to extremize the action. A solution of the saddle point equations was obtained more than twenty years ago by Bray and Moore (BM) \cite{BMan} and another solution was proposed in \cite{PP}. The two solutions lead to a very different complexity, see {e.g.} \cite{noian}, and one must decide which is the correct one. 
The action is invariant under the so called Becchi-Rouet-Stora-Tyutin (BRST) transformation \cite{BRST,ZZ}. This transformation mixes bosonic and fermionic variables {\it i.e.} it is a SUSY transformation.
An important difference between the BM solution and the second solution is that the BM solution breaks the SUSY of the action, as firstly pointed out by Kurchan in \cite{K91}, while the second solution preserves the SUSY \cite{CGPM}.

Recently it has been shown \cite{noiquen} that the SUSY solution is inconsistent (except at the lower-band edge, see \cite{BMY,noiquen}) in full replica-symmetry-breaking (FRSB) models like the SK  model (at variance with 1RSB models), thus the BM solution remains the only viable candidate to describe a finite complexity in these models and the knowledge of the consequences of SUSY breaking becomes very important.

As a first consequence of BRST invariance, one finds that the  
the prefactor of the exponential contribution to the complexity vanishes when evaluated at a non-SUSY saddle point, like the BM one.
This result was obtained by Kurchan in \cite{K91} for the total number of solutions, here we further extend it to solutions at any  given value of the free energy. 
The result is obtained in two steps. In the next two sections we start recalling that the expression of $\Sigma(f)$ (the logarithm of the number of TAP solutions of a given free energy $f$) is invariant under a  microscopic supersymmetric transformation that is a generalization of the standard BRST transformation \cite{juanpe}.
After performing the disorder averages the integral over the microscopic action is expressed as an integral over a macroscopic action depending on eight macroscopic variables (four fermionic and four bosonic), that are introduced through standard manipulation such as even and odd Hubbard-Stratonovich transformations. Following \cite{K91} we note that this action too is invariant under a macroscopic SUSY transformation (that depends on the parameter $u$ conjugated to the free energy $f$) between its eight variables. 
The  second step consists in applying a general result that states that if the action is invariant under a SUSY transformation then the prefactor of the exponential contribution is zero at all orders in an expansion in powers of $1/N$. This result is completely general and was obtained by Kurchan \cite{K91} who applied it to the total number of solutions. Since we have shown that the macroscopic action at any give value of the free energy is also BRST invariant we can apply it in this case too. For completeness Kurchan argument is reported in section \ref{sec:thepref2}.

In section \ref{sec:thepref1} we provide a new simple argument to derive the vanishing of the $O(1)$ part of the prefactor. According to it the $O(1)$ part of the prefactor vanishes because it is proportional to the determinant of the Hessian of the fermionic variables which must be zero due to SUSY breaking. 
This argument is very intuitive and illustrates the similarities between the breaking of a standard bosonic symmetry, that produces Goldstone modes and infinite fluctuations, and the breaking of a SUSY, that produces a zero prefactor.
The above theoretical predictions on the vanishing of the prefactor on the BM solution are also verified numerically at the end of section \ref{sec:themacroaction}

Recently Aspelmaier, Bray and Moore \cite{ABM} have shown that the spectrum of the Hessian of the TAP free energy contains an isolated eigenvalue besides the continuous band. 
They found numerically that this eigenvalue vanishes on the BM solution.
The presence of a zero  eigenvalue in the BM solution is very important since it is a possible explanation for the vanishing of the prefactor and for the apparent violation of the Morse theorem \cite{K91, CGPM, noian}, see discussion in the conclusions.
In section \ref{sec:theiseigen} we prove rigorously the vanishing of the isolated eigenvalue showing that it is another consequence of SUSY breaking. Furthermore we discuss the behavior of the eigenvector corresponding to the isolated eigenvalue at the lower band edge of the complexity where the continuous band of eigenvalues extends down to zero.
At the end we briefly discuss the results. 

\section{Complexity and  Supersymmetry}
\label{sec:compesup}

We are interested in the solutions of the TAP equations $\partial_i F_{TAP}(m)=0$, where $F_{TAP}(m)$ is  a given model-dependent TAP free energy.
In particular we want to compute the density of TAP solutions of a given free energy,
\beq
\rho(f)=\sum_{\alpha=1}^{\cal N} \delta[F_{TAP}(m^\alpha)-Nf] \ ,
\label{rho}
\eeq
In the previous expression TAP solutions are labeled by the index $\alpha$, and $\{m^\alpha\}$
indicates the corresponding set of local magnetizations.
The density can be expressed as an integral over the whole $m$-space
of a delta function of the TAP equations:
\begin{eqnarray}\rho(f) & = &\sum_{\alpha=1}^{\cal N} \int \prod_i dm_i\ \delta(m_i-m_i^\alpha) \ \delta[F_{TAP}(m^\alpha)-Nf] =
\non
\\
& = & \int \prod_i dm_i\ \delta(\partial_i F_{TAP}(m)) \ |\det (\partial_i \partial_j F_{TAP}(m))| \ \delta[F_{TAP}(m)-Nf] 
\ ,
\label{ghirri}
\end{eqnarray}
In order to use an exponential representation of the determinant its modulus is dropped. This is a crucial step that corresponds to count each solution with a weight proportional to the sign of its determinant. The resulting object is not the number of solutions with free energy $f$ but rather the number of solutions with positive determinant minus the number of solutions with negative determinant. 
Therefore, as we said in the introduction, the problem splits in two: from one hand we compute the sum without the modulus but on the other hand we want to obtain information on the actual the complexity, {\it i.e.} on the sum {\it with} the modulus.
In particular we want to compute the total complexity but we want also to recover the trivial result that the sum without the modulus of all solutions be equal to $1$ because of the Morse theorem {\it i.e.} $\int df \rho (f)=1$.
We also use an exponential representation for the delta function,
\beqa
 \prod_i \delta(\partial_i F_{TAP}) &=& \int_{-i\infty}^{+i\infty} \prod_i \frac{dx_i}{2\pi i}
\ \exp\left(\sum_i x_i \partial_i F_{TAP}(m)\right) \ ,\\
\det (\partial_i \partial_j F_{TAP}) &=& \int_{-\infty}^{+\infty} \prod_i d\bar\psi_i\, d\psi_i 
\ \exp\left[\sum_{ij} \bar\psi_i  \psi_j  \partial_i \partial_j F_{TAP}(m) \right] \ ,
\eeqa
where $\{\bar\psi, \psi\}$ are anti-commuting Grassmann variables. 
The delta function over the free energy is also expressed as an exponential
\beq
\delta(F_{TAP}-N\, f)  = \int_{-i\infty}^{+i\infty} \frac{du}{2\pi i} 
\ \exp[\,u\,(F_{TAP}-Nf)] \ ,
\eeq
In this way we can write \cite{ZZ},
\beq
\rho(f)=\int du{\cal D}m\, {\cal D}x \,{\cal D}\bar\psi\, {\cal D}\psi\ \
e^{S(m,x,\bar\psi,\psi,u)} \ ,
\label{int2}
\eeq
where the action $S$ is given by,
\beq
S(m,x,\bar\psi,\psi,u)= \sum_i x_i \partial_i F_{TAP}(m) + \sum_{ij}\bar\psi_i  \psi_j  
\partial_i \partial_j F_{TAP}(m) + u[F_{TAP}(m)] -u f\ .
\label{action}
\eeq
A key property of action (\ref{action}) is its invariance under a generalization \cite{PaSou1,PaSou2,juanpe} of the 
Becchi-Rouet-Stora-Tyutin (BRST) supersymmetry \cite{BRST,ZZ}:
if $\epsilon$ is an infinitesimal Grassmann parameter, it is easy to verify that the action  
(\ref{action}) is invariant under the following transformation,
\beq
\delta m_i = \epsilon\, \psi_i \quad\quad
\delta x_i = -\epsilon \, u \, \psi_i \quad\quad
\delta \bar\psi_i = -\epsilon\, x_i \quad\quad
\label{brst}
\eeq
The TAP free energy of the SK model is given by \cite{TAP},
\beq
F_{TAP}(m)=-\frac{1}{2} \sum_{ij} J_{ij} m_i m_j + \frac{1}{\beta} \sum_i \phi_0(q,m_i) \ ,
\label{ftap}
\eeq
with,
\beqa
\phi_0(q,m)&=&
\frac{1}{2}(1+m)\, \log\left[\frac{1}{2}(1+m)\right] +
\frac{1}{2}(1-m)\, \log\left[\frac{1}{2}(1-m)\right] -\frac{\beta^2}{4}(1-q)^2
\non
\\
&=& \frac{1}{2}\log(1-m^2) + m\,\tanh^{-1}(m) -\log 2 -\frac{\beta^2}{4}(1-q)^2 \ .
\label{phi0}
\eeqa
The variables $m_i$ are the local magnetizations, and $q$ is  the self-overlap of the TAP states,
\beq
q=\frac{1}{N}\sum_i m_i^2 \ ,
\eeq
while the quenched couplings $J_{ij}$ are Gaussian random variables with zero mean and variance $1/N$.
The TAP equations and the Hessian of the free energy are respectively,
\beqa
\beta \, \partial_i F_{TAP}(m) &=& -\beta \sum_{j\neq i} J_{ij} m_j + \phi_1(q,m_i) = 0 \ , 
\label{tapeque}
\\
\beta \, \partial_i \partial_j F_{TAP}(m) &=& -\beta J_{ij} + \phi_2(q,m_i) \, \delta_{ij} -{2 \beta^2 \over N}m_i m_j \ ,
\label{taphess}
\eeqa
with
\beqa
\phi_1(q,m) &=& \beta^2 (1-q) m + \tanh^{-1}(m) \ , \\
\phi_2(q,m) &=& \beta^2 (1-q) + \frac{1}{1-m^2} \  \ .
\label{phi}
\eeqa
The last term of order $O(1/N)$ in (\ref{taphess}) is very important for the present discussion. 
We perform an {\it annealed} calculation of the number of TAP states, 
i.e. we directly average $\rho(f)$ in (\ref{rho})
over the distribution of the quenched couplings $J_{ij}$. 
Thus, the average number of TAP states becomes,
\beq
\overline{\rho(f)}= \int {\cal D}J \, P(J) \ {\cal D}m\, {\cal D}x \,{\cal D}\bar\psi\, {\cal D}\psi\, du \
e^{\beta S(m,x,\bar\psi,\psi,u)} \ =\int \ {\cal D}m\, {\cal D}x \,{\cal D}\bar\psi\, {\cal D}\psi\, du \
e^{\beta S_{\rm av}(m,x,\bar\psi,\psi,u)} ,
\label{pino}
\eeq
where
\begin{equation}
\beta S_{\rm av}=\ln \int {\cal D}J \, P(J) 
e^{\beta S(m,x,\bar\psi,\psi,u)} \ .
\label{pino2}
\end{equation}

\section{The Macroscopic Action}
\label{sec:themacroaction}

The term in the action (\ref{action}) that depends on $J_{ij}$ is
\begin{equation}
\beta S_J=-\beta \sum_{i<j} J_{ij}(x_im_j+x_jm_i+\overline{\psi}_j \psi_i+\overline{\psi}_i \psi_j+u m_im_j)
\end{equation}
By averaging the exponential of $S_J$ with respect to the $J$'s the contribution of the previous term to the averaged action $S_{\rm av}$ turns out to be:
\begin{displaymath}
{\beta^2  \over 2N}\sum_{i<j}(x_im_j+x_jm_i+\overline{\psi}_j \psi_i+\overline{\psi}_i \psi_j+u m_im_j)^2=
\end{displaymath}
\begin{equation}
{\beta^2  \over 4N}\sum_{ij}(x_im_j+x_jm_i+\overline{\psi}_j \psi_i+\overline{\psi}_i \psi_j+u m_im_j)^2-{\beta^2  \over 4N}\sum_{i}(2x_im_i+2\overline{\psi}_i \psi_i+u m_im_i)^2
\label{1p2}
\end{equation}
Introducing the following shorthand notation for various macroscopic quantities,
\begin{equation}
\begin{array}{cccc}
R={1 \over N}\sum_i m_i x_i;  &  T={1 \over N}\sum_i \overline{\psi}_i \psi_i; & q={1 \over N}\sum_i m_i m_i;  &
\\
\theta={1 \over N}\sum_i \psi_i m_i; &  \overline{\theta}={1 \over N}\sum_i \overline{\psi}_i m_i; & \nu ={1 \over N}\sum_i \psi_i x_i; &\overline{\nu} ={1 \over N}\sum_i \overline{\psi}_i x_i;   
\end{array}
\end{equation}
we can express the first term in eq. (\ref{1p2})
as
\begin{equation}
N \beta^2\left(  {u^2 q^2 \over 4} + u  q R+{q\over 2 N}\sum_ix_i^2+{ R^2 \over 2}+ u \overline{\theta}\theta+\overline{\theta}\nu +\overline{\nu}\theta-{T^2\over 2}\right)
\end{equation}
This term is order $N$ while the second term in eq. (\ref{1p2}) is order $1$ since it is the sum divided by $N$ of $N$ terms that are local in $i$:
\begin{equation}
-{\beta^2  \over 4N}\sum_{i}(2x_im_i+2\overline{\psi}_i \psi_i+u m_im_i)^2={1 \over N}\sum_i F_{loc}(m_i,x_i,\overline{\psi}_i,\psi_i)
\label{floc}
\end{equation}
Looking at (\ref{action}), we see that the averaged action (\ref{pino2}) is given by
\begin{eqnarray}
\beta S_{\rm av} & = & \sum_ix_i \phi (q,m_i)+\sum_i\overline{\psi}_i\psi_i \phi_2(q,m_i)+u \sum_i\phi_0(q,m_i)-N \beta u f+
\nonumber
\\ 
& + & N \beta^2\left(  {u^2 q^2 \over 4} + u  q R+{q\over 2 N}\sum_ix_i^2+{ R^2 \over 2}+ u \overline{\theta}\theta+\overline{\theta}\nu +\overline{\nu}\theta-{T^2\over 2}\right)+
\nonumber
\\
 & + &{1 \over N}\sum_i F_{loc}(m_i,x_i,\overline{\psi}_i,\psi_i)-2 \beta^2 N \overline{\theta} \theta 
\label{Sav}
\end{eqnarray}
The last term comes from the term of order $O(1/N)$ in the Hessian (\ref{taphess}). Indeed from (\ref{action}) we see that the contribution to $S_{\rm av}$ of this term  is  
\begin{equation}
-{2 \beta^2 \over N}\sum_{ij}\overline{\psi}_j\psi_i m_j m_i =-2 N\beta^2 \overline{\theta}\theta
\end{equation}
Note that the action (\ref{Sav}) is still invariant under the BRST transformation
\begin{equation}
\delta m_i= \epsilon \psi_i \ ;\ \ \delta x_i=-\epsilon u \psi_i \ ;\ \ \ \delta \overline{\psi}_i=-\epsilon x_i \ ;
\end{equation}
To proceed further we eliminate the squares of the macroscopic
quantities $R$,$T$,$q$, $\overline{\theta}$, $\theta$,
$\overline{\nu}$, $\nu$, in the action (\ref{Sav}). In particular the
exponentials $\exp[R^2]$ and $\exp[-T^2]$ are expressed as gaussian
integrals over respectively the real and the complex axes, introducing the variable $r$ and $t$. We also use an integral
representation of a delta function on $q$ that introduces an extra
parameter $\lambda$ conjugated with $q$. The fermionic part is
simplified through the following odd Hubbard-Stratonovich
transformation  that introduces four new parameters  $\{\overline{\rho},\rho,\overline{\mu},\mu \}$
\begin{displaymath}
\exp[-\left(\begin{array}{cc}
\overline{\theta}
&
\overline{\nu}
\end{array}\right)
\bf{L}
\left(\begin{array}{c}
\theta \\ \nu
\end{array}\right) ]={\rm det}{\bf L}\int  d\rho d\mu d\overline{\rho}d\overline{\mu} \exp[\left(\begin{array}{cc}
\overline{\rho}
&
\overline{\mu}
\end{array}\right)
\bf{L^{-1}}
\left(\begin{array}{c}
\rho \\ \mu
\end{array}\right)+
\end{displaymath} 
\begin{equation}
+\left(\begin{array}{cc}
\overline{\theta}
&
\overline{\nu}
\end{array}\right)\left(\begin{array}{c}
\rho \\ \mu
\end{array}\right) +\left(\begin{array}{cc}
\overline{\rho}
&
\overline{\mu}
\end{array}\right)
\left(\begin{array}{c}
\theta \\ \nu
\end{array}\right) ]
\label{Hdis} 
\end{equation}
Where according to eq. (\ref{Sav}) the matrix {\bf L} is given by :
\begin{equation}
{\bf L}=-\beta^2\left(\begin{array}{cc}
u-2 & 1
\\
1 & 0
\end{array}\right)
\label{Beq}
\end{equation}
By means of the various transformations the total action can be expressed as an integral over the variable $u$ and eight macroscopic bosonic and fermionic variables $\Theta$ $\equiv$ $\{r$,$t$,$q$,$\lambda$,$\overline{\rho}$,$\rho$,$\overline{\mu}$,$\mu \}$
\begin{equation}
\rho_s=\int du \ d \Theta \exp[N \Sigma_1+N \Sigma_2 ]
\label{ess}
\end{equation}
Where 
\begin{eqnarray}
\Sigma_1 & = & -\beta u f-\lambda q+q r u -{r^2 \over 2 \beta^2}+{t^2 \over 2
 \beta^2}-{u \beta^2\over 4}+{q u \beta^2\over 2}-{q^2 u \beta^2\over
 4}+{q^2 u^2 \beta^2\over 4}
\nonumber
\\
& +& \overline{\mu}\mu+(2-u)\overline{\mu}\rho+\overline{\rho}\rho
\end{eqnarray}
and
\begin{eqnarray}
\Sigma_2 &=&\log\left[\int d m d x d \psi d \overline{\psi}\
   \exp \Bigl[ x
   \phi(q,m)+\overline{\psi}\psi\phi_2(q,m)+u \phi_0(q,m)+ \right.
\nonumber
\\
& + & {q \beta^2  x^2 \over 2} +{1 \over N}
   F_{loc}(m,x,\psi,\overline{\psi})+rmx+t\overline{\psi}\psi+\lambda
   m^2+{u\beta^2\over 4}(1-q)^2+
\nonumber
\\
& - &\left. \mu \beta m \psi
   -\overline{\psi}m\overline{\rho}\beta-\rho \beta x \psi
   -\overline{\psi}x \overline{\mu}\beta  \Bigr] \right]
\label{S2}
\end{eqnarray}
Following Kurchan \cite{K91} we note that the original invariance of the action under the BRST transformation
reflects itself in the fact the 
macroscopic action $\Sigma_1+\Sigma_2$ is invariant under the following SUSY
transformation:
\begin{equation}
\left\{
\begin{array}{ccc}
\delta \mu & = & \left( -{r u \over \beta }-u \beta + q u \beta + {2
 \lambda \over \beta}  \right)\epsilon
\\
\delta \rho & = & \left( {r  \over \beta }-{t \over \beta} -q u \beta   \right)\epsilon
\\
\delta \overline{\mu} & = & 0
\\
\delta \overline{\rho} & = & 0 
\\
\delta q & = & -{2 \overline{\mu} \over \beta}\epsilon
\\
\delta \lambda  & = & 0
\\
\delta r & = & -(2 \beta \overline{\mu}-\beta \overline{\rho})\epsilon 
\\
\delta t & = & ((u-2) \beta \overline{ \mu }-\beta \overline{\rho})\epsilon
\end{array}
\right. 
\label{macro}
\end{equation}
To obtain the previous transformation we can consider a generical SUSY transformation
 and determine its parameter in order that $\Sigma_2$ be invariant. To do this we must use the fact that $\Sigma_2$ is  invariant if we transform the integrand in (\ref{S2}) with a transformation like (\ref{brst}). This happens because of the vanishing of the surface terms at $m=\pm 1$ and $x =\pm i\infty$, as can be easily checked. Thus the variation of $\Sigma_2$ induced by the transformation (\ref{brst}) on its integrand is a quantity, depending on the macroscopic variables, that must sum up to zero and can be added to the variation induced by a generical SUSY transformation between the macroscopic variable in order to determine  (\ref{macro}).

\section{The Prefactor of the Exponential (I)}
\label{sec:thepref1}

In this section we discuss the general problem of evaluating a SUSY action by means of the saddle point method.
We are interested in object of this kind:
\begin{equation}
\int d \{ {\Phi} \}d \{ { \Psi} \}\exp[N F( \Phi , \Psi )]
\label{int}
\end{equation}
Where $\{ {\Phi} \}$ and $\{ {\Psi} \}$ are two sets of respectively bosonic and fermionic variables.
Expanding $F$ around a saddle point with the substitution
$\Phi=\Phi_0+\delta \Phi/ \sqrt{N}$ and $\Psi=\Psi_0+\delta \Psi/ \sqrt{N}$ we obtain
\begin{equation} 
N F( \Phi , \Psi )=N F( \Phi_0,\Psi_0)+{\partial^2  F\over \partial
 \Phi \partial \Phi }\delta \Phi \delta \Phi+ {\partial^2  F\over \partial
 \overline{\Psi} \partial \Psi }\delta \overline{\Psi} \delta \Psi +O({1\over \sqrt{N}})  \ ,
\end{equation}
and the integral (\ref{int}) over the action can be  rewritten as
\begin{equation}
\exp[N F( \Phi_0,\Psi_0] \int d \{ \Phi \}d \{ \Psi \}\exp[{\partial^2  F\over \partial \Phi \partial \Phi }\delta \Phi \delta \Phi+ {\partial^2  F\over \partial
 \overline{\Psi} \partial \Psi }\delta \overline{\Psi} \delta \Psi +O({1\over \sqrt{N}})]
\end{equation}
The second factor in the above expression gives the corrections to the leading exponential
contribution of the saddle point.
We note that because of their very nature the integral on the fermionic variables in (\ref{int}) in any case does not give a term of the form $e^N$; this feature can be accounted for by saying that the fermions must be ``set to zero" on the saddle point, that is we always have $\Psi_0=0$.   

In the following section we will report a general argument \cite{K91} in order to prove that if the action $F$ possess a SUSY the prefactor of the exponential contribution on a non-SUSY saddle point is zero at all order in an expansion in powers of $1/N$.
In this section instead we provide a simple argument to prove the vanishing of the first term in the expansion.
According to the previous expression this term is simply given by:
\begin{equation}
\int d \{ \Phi \}d \{ \Psi \}\exp[{\partial^2  F\over \partial \Phi \partial \Phi }\delta \Phi \delta \Phi+ {\partial^2  F\over \partial
 \overline{\Psi} \partial \Psi }\delta \overline{\Psi} \delta \Psi]=\det\left( {\partial^2  F\over \partial
 \overline{\Psi} \partial \Psi } \right)\det \left( {\partial^2  F\over \partial
 \Phi \partial \Phi } \right) ^{-1/2}
\label{detfe}
\end{equation}
The invariance under the BRST transformation reads 
\begin{equation}
{\mathbf \Phi} \equiv \{\Phi,\Psi\} \ ;\ \     0=DF=\delta {{\mathbf \Phi}}_i {\partial F \over \partial {{\mathbf \Phi}}_i} \ \
\ {\rm for}\ \ \  \delta {\mathbf \Phi}_i=T_{ik}{\mathbf \Phi}_k \epsilon
\end{equation}
where $ {{\mathbf \Phi}}$ represents the total set of fermionic and
bosonic fields.
Deriving the previous expression with respect to a generic component
of  $ {{\mathbf \Phi}}$
we have
\begin{equation}
{\partial F \over \partial {\mathbf \Phi}_k \partial
 {\mathbf \Phi}_m}T_{kl}{\mathbf \Phi}_l+{\partial F\over \partial {\mathbf \Phi}_k} T_{km}=0\label{jac0}
\label{vec}
\end{equation}
On a saddle point  the second term in the l.h.s. of the previous equation is zero by definition therefore, if $T_{kl}{\mathbf \Phi}_l$ is different from zero,
it is an eigenvector of the matrix $\partial F / \partial {\mathbf \Phi}_k \partial {\mathbf \Phi}_m$ with  zero eigenvalue.
The vector $T_{kl}{\mathbf \Phi}_l$ is strictly zero by definition on a SUSY saddle point while it is non-zero on a non-SUSY saddle point. The fermionic component of ${\mathbf \Phi}$, i.e. $\{\overline{\mu},\mu,\overline{\rho},\rho\}$ are ``set to zero" on both the SUSY and the non-SUSY saddle point, as a consequence the bosonic component of the variation vector $T_{kl}{\mathbf \Phi}_l$ is strictly zero on both saddle point. Thus only the fermionic  component of $T_{kl}{\mathbf \Phi}_l$ on the non-SUSY saddle point can be non-zero, this means that  the part of the Hessian with a zero eigenvalue is the fermionic one, as a consequence:
\begin{equation}
\det\left( {\partial^2  F\over \partial
 \overline{\Psi} \partial \Psi } \right)_{\rm NO-SUSY}=0 
\end{equation}
Coming back to equation (\ref{detfe}) we see that the $O(1)$ contribution to the prefactor of the exponential term vanishes on a non-SUSY saddle point. The same argument does not apply to a SUSY saddle point where the $T_{il}{\mathbf \Phi}_l$ is strictly zero and equation (\ref{jac0}) is verified trivially.

In order to verify directly the previous argument we have computed the fermionic determinant expanding $F=\Sigma_1+\Sigma_2$ in powers of $\{ \overline{\Psi},\Psi\}\equiv \{\overline{\mu},\mu\,\overline{\rho},\rho\}$.
The resulting expression is 
\begin{equation}
{\partial^2  F\over \partial
 \overline{\Psi} \partial \Psi }=
\left(\begin{array}{cc}
1+\beta^2 \langle \overline{\psi}\psi m x \rangle & 
\beta^2 \langle\overline{\psi}\psi m^2 \rangle
\\
2-u+\beta^2 \langle\overline{\psi}\psi x^2 \rangle & 
1+\beta^2 \langle \overline{\psi}\psi m x \rangle
\end{array}\right)
\end{equation}
Where the square brackets mean averages computed with respect to the integrand of $\Sigma_2$ (equation \ref{S2}) with the fermionic variables  $\{\overline{\mu},\mu\,\overline{\rho},\rho\}$ set to zero. Performing the integration over $\overline{\psi},\psi$ and over $x$ we obtain the components of the Fermionic Hessian as averages with respect to the variable $m$. Skipping from the macroscopic variable $\{r,t,\lambda,q,u\}$ to the BM variables $\{B,\Delta,\lambda,q,u\}$ of  \cite{BMan} we obtain:
\begin{equation}
{\partial^2  F\over \partial
 \overline{\Psi} \partial \Psi }=
\left(\begin{array}{cc}
1+\beta^2 \langle (1-m^2)m{m \Delta -\tanh^{-1}  m\over q \beta^2}                         \rangle 
& 
\beta^2 \langle (1-m^2) m^2 \rangle
\\
2-u+\beta^2 \langle (1-m^2){(\tanh^{-1}  m-m \Delta)^2-q \beta^2  \over q^2 \beta^4}                \rangle 
& 
1+\beta^2 \langle (1-m^2)m{m \Delta -\tanh^{-1} m\over q \beta^2}                         \rangle 
\end{array}\right)
\label{det}
\end{equation}
Where the square brakets represent averages computed with respect to the standard BM action. 
We checked numerically that when computed on the BM solution the determinant of the previous matrix is zero.

\section{The Prefactor of the Exponential (II)}
\label{sec:thepref2}
In this section we briefly recall the argument of Kurchan \cite{K91} in order
to show that the prefactor vanishes at all orders in the expansion in
powers of $1/N$.
In the global variables $\{{\mathbf \Phi}\}$ of the previous section the integral over the action reads:
\begin{equation}
e^{N F({\mathbf \Phi}_0)}\int d \{{\mathbf \Phi}\}\exp[F_{2}{\mathbf \Phi} {\mathbf \Phi}+{1 \over
  \sqrt{N}}F_3{\mathbf \Phi} {\mathbf \Phi} {\mathbf \Phi} +{1 \over N}F_4 {\mathbf \Phi} {\mathbf \Phi} {\mathbf \Phi} {\mathbf \Phi}+\dots] \ ,
\label{acgen}
\end{equation}
and the expansion in powers of $1/\sqrt{N}$ is:
\begin{equation}
e^{N F({\mathbf \Phi}_0)}\int d \{{\mathbf \Phi}\}\sum_{r=0}^\infty \left({1 \over
 \sqrt{N}} \right)^r A_r({\mathbf \Phi}) \ .
\label{prefex}
\end{equation}
Now we make the change of variables ${\mathbf \Phi}={\mathbf \Phi}_0+{\mathbf \Phi}'/ \sqrt{N}$. The integrand must be invariant under the BRST transformation $\delta {\mathbf \Phi}_i=\epsilon \, T_{ik}{\mathbf \Phi}_k$, that in
the new variables reads
\begin{eqnarray}
D & = & T_{ik}{\mathbf \Phi}_k {\partial \over \partial {\mathbf \Phi}_i}=T_{ik}{\mathbf \Phi}_{k}^{(0)}
{\partial \over \partial {\mathbf \Phi}_i}+{1\over
 \sqrt{N}}T_{ik}{\mathbf \Phi}_k'{\partial \over \partial {\mathbf \Phi}_i}=
\nonumber
\\
& = & \sqrt{N}T_{ik}{\mathbf \Phi}_k^{(0)}{\partial \over \partial {\mathbf \Phi}_i'}+T_{ik}{\mathbf \Phi}_k' {\partial \over \partial {\mathbf \Phi}_i'}=\sqrt{N}D_0+D_1
\label{defD}
\end{eqnarray}
The invariance reads
\begin{equation}
D \sum_{r=0}^\infty \left({1 \over
 \sqrt{N}} \right)^r A_r({\mathbf \Phi}')=\sqrt{N}D_0 A_0+\sum_{r=0}^\infty
\left( {1 \over \sqrt{N}} \right)^r (D_0A_{r+1}+D_1 A_r)=0
\end{equation}
Since it must be satisfied for each value of $N$ each term in the
r.h.s must vanish separately leading to:
\begin{equation}
D_0A_0=0;   \ \ \ \ \ \ D_0A_{r+1}+D_1 A_r=0
\end{equation}
The operators $D_0$ and $D_1$ satisfy
\begin{equation}
D_0D_1+D_1D_0=0  \ \ \ D_0^2=0 \ \ \ D_1^2=0
\end{equation}
Note that the equation $D_0A_0=0$ is trivially satisfied on a BRST-susy
saddle point, indeed according to eq. (\ref{defD}) $D_0$ depends on
the vector of the BRST transformation $T_{ik}{\mathbf \Phi}^{0}_k$ that is zero by
definition on a symmetric saddle point leading to $D_0=0$.
Instead if $D_0\neq 0$ we may write  $D_0 B=1$ with $B$  a constant
times one of the fermionic variables  ${\mathbf \Phi}'_i$ for which
$T_{ik}{\mathbf \Phi}'_k$ is non-zero.
Multiplying by $(D_0 B)^k=1$ and integrating repeatedly by parts we have 
\begin{eqnarray}
\int d \{{\mathbf \Phi}'\} A_r({\mathbf \Phi}') & = & \int d \{{\mathbf \Phi}'\} (D_0B)^{r+1}A_r({\mathbf \Phi}')
=
\nonumber 
\\
& = & \int d \{{\mathbf \Phi}'\} (D_0B)^{r}A_{r-1} (D_1B)=
\nonumber
\\
& = & \dots = \int d \{{\mathbf \Phi}'\} D_0BA_{0} (D_1B)^r=
\nonumber
\\
& = & \int d \{{\mathbf \Phi}'\} BD_0A_{0} (D_1B)^r=0
\label{kua}
\end{eqnarray}
Thus looking to (\ref{prefex}) and (\ref{kua}) we see that on a non-SUSY saddle point the prefactor of the exponential vanishes at all order in powers of $1/N$. As noted in \cite{K91} this may happens either if the prefactor is strictly zero at all orders or if it is of the form $e^{-a N}$  with $a>0$.

Let us note that the action (\ref{ess}) is not precisely of the form (\ref{int}) since $\Sigma_2$ defined in (\ref{S2}) contains an extra dependence on $1/N$ coming from the local terms (\ref{floc}).
These terms obviously enter the $1/N$ expansion but they do not change the result on the vanishing of the expansion at all orders. Indeed the presence of these extra terms in (\ref{acgen}) does not change the form (\ref{prefex}) and the subsequent derivation.
The same is true for the argument of the previous section, indeed these terms give a multiplicative correction different from one to the prefactor that cannot change the final result that depends on the vanishing of the fermionic determinant.
However it is important to note that these terms should be considered in the construction of $1/N$ corrections to the real complexity, {\it i.e.} the one with the modulus.

\section{The Isolated Eigenvalue}
\label{sec:theiseigen}

The density of eigenvalues corresponding to the Hessian without the last term in (\ref{taphess})  can be obtained through standard theorems on random matrices (see \cite{BMsp,Ple1}). It  turns out to have always a positive support with a lower band edge going to zero if the quantity 
\begin{displaymath}
x_p=1-\beta^2\sum_i(1-m_i^2)^2
\end{displaymath}
goes to zero. The BM solution has a finite value of $x_P$ meaning that the corresponding TAP solutions has a definite positive Hessian. On the other hand the Morse theorem predicts that there is an equal number  of solutions with positive and negative determinant of the Hessian, so the problem is: where are these negative determinant solutions?
Note that the answer cannot come from the finite-$N$ tail of the eigenvalue density if we assume
 that at finite $N$ the eigenvalues outside the support are concentrated at a distance $N^{-1/6}$ from the lower band edge, as in most cases in random matrix theory \cite{mehta}. 

In \cite{ABM} Aspelmaier, Bray and Moore (ABM) has shown that
the inclusion of  the last term in the Hessian (\ref{taphess}) modifies the continuous spectrum of the first two terms with the addition of an isolated eigenvalue.
Most importantly they checked numerically that this isolated eigenvalue vanishes on the BM solution solving the apparent inconsistency with the Morse theorem.
In this section we prove rigorously the vanishing of the isolated eigenvalue showing that it is a consequence of SUSY breaking.

The appearance of the isolated eigenvalue is connected to the fact that the last term in the Hessian (\ref{taphess}) is proportional to a projector.
Given two symmetric matrices $A$ and $B$ that differ by a projector $P=|\alpha \rangle \langle \alpha |$ we have:
\begin{equation}
A= B + P \longrightarrow \det A= (1+\langle \alpha |B^{-1}|\alpha \rangle  )\det B \label{abp}
\end{equation}

This relation can be easily proved considering an orthonormal base with, say, the first element equal to $|\alpha \rangle $, such that the only non-zero component of $P_{ij}$ is $P_{11}=1$.
The same argument applies to $B=A-P$ leading to:
\begin{equation}
1+\langle \alpha |B^{-1}|\alpha \rangle  ={1 \over 1-\langle \alpha |A^{-1}|\alpha \rangle  }
\label{ABinv}
\end{equation}
Thus the projector modifies the spectrum, in particular if $(1+\langle \alpha |B^{-1}|\alpha \rangle  )$ is zero the isolated eigenvalue vanishes and the corresponding eigenvector is $B^{-1}|\alpha\rangle$.

The quantity $(1+\langle \alpha |B^{-1}|\alpha \rangle )$ is essentially the ABM quantity $1-2 \beta^2 H $ that must be zero in order for the isolated eigenvalue to vanish \cite{ABM}.
By direct inspection one can check that the expression for the determinant of  the fermionic Hessian eq. (\ref{det}) turns out to be proportional to $1-2 \beta^2 H$, see equation (8-11)
in \cite{ABM}. 
Since in section \ref{sec:thepref1} we proved that the fermionic determinant must vanish on a non-SUSY saddle point like the BM one this also prove rigorously that the isolated eigenvalue vanishes:
\begin{equation}
\left. 1-2 \beta^2 H \right|_{\rm NO-SUSY} \propto  \det\left( {\partial^2  F\over \partial
 \overline{\Psi} \partial \Psi } \right)_{\rm NO-SUSY}=0   \ .
\end{equation}
In the following we prove directly the connection between the isolated eigenvalue and the determinant of the fermions.
The relation (\ref{ABinv}) allows to consider the quantity $1-\langle \alpha |A^{-1}|\alpha \rangle  $ instead of $(1+\langle \alpha |B^{-1}|\alpha \rangle )$.
That is we can use the full Hessian 
\begin{equation}
A_{ij}=\partial^2F/\partial m_i \partial m_j \ ,
\end{equation} 
instead of the Hessian without the projector term.
Therefore we must evaluate $\langle m |A^{-1}|m \rangle /N$ on all solutions of the TAP equations:
\begin{equation} 
\frac{\sum_{\alpha} {1 \over N} \sum_{ij} m_i \left( {\partial^2F \over \partial m \partial m} \right)^{-1}_{ij}m_j }{\sum_{\alpha} }=\frac{\int {1 \over N} \sum_{ij} m_i \psi_i \overline{\psi}_j m_j \exp[S[m,x,\psi, \overline{\psi}]]}{\int \exp[S[m,x,\psi, \overline{\psi}]]}=\frac{\int N \theta \overline {\theta} \exp[\beta S]}{\int \exp[\beta S]}
\label{tete}
\end{equation}
Where the action $\beta S$ is given by eq. (\ref{Sav}). The presence of the term $\theta \overline {\theta} $ in the integrand must be taken into account when performing the decoupling of the variables $\{ \theta , \overline{\theta} , \nu , \overline{\nu}\}$. In general this can be achieved by multiplying the integrand in the l.h.s. of eq. (\ref{Hdis}) for some proper prefactor depending on the matrix {\bf L} and on the integration variables $\{\mu , \overline{\mu} ,\rho,\overline{\rho}\}$. In our case {\bf L} is given by eq. (\ref{Beq}) and the prefactor turns out to be $\mu \overline{\mu}$. Thus  we obtain that:
\begin{equation}
{1 \over N}\langle m |A^{-1}|m \rangle  \propto \frac{\int d \Theta N \mu \overline{\mu}\exp[N \Sigma_1+N \Sigma_2 ]
}{\int d \Theta \exp[N \Sigma_1+N \Sigma_2 ]
}
\label{relan}
\end{equation}
By computing the previous quantity at the saddle point we obtain that the bosonic contributions of order $e^{N \Sigma}$ and of order $1$ (given by the bosonic determinant, see eq. (\ref{detfe})) are equal in the numerator and in the denominator. Instead the fermionic contributions are different. In the denominator the contribution is given by the fermionic determinant as in (\ref{detfe}) while in the numerator the presence in the integrand of the factor $\mu \overline{\mu}$ gives only  a contribution equal to $\partial^2 F / \partial \rho \partial \overline{\rho}$ that is a finite quantity proportional to the diagonal terms of eq. (\ref{det}).
Thus we have that:
\begin{equation}
{1 \over N}\langle m |A^{-1}|m \rangle  \propto  {\partial^2 F \over \partial \rho \partial \overline{\rho}} \det\left( {\partial^2  F\over \partial
 \overline{\Psi} \partial \Psi } \right)^{-1}\longrightarrow 1+\langle \alpha |B^{-1}|\alpha \rangle  \propto \det\left( {\partial^2  F\over \partial
 \overline{\Psi} \partial \Psi } \right)
\end{equation}
Going back to equation (\ref{abp}) we see that
 the determinant of the Hessian is proportional to the determinant of the fermionic components of the macroscopic action and vanishes on the non-SUSY solutions.

According to the prediction of the BM solution the determinant of the Hessian TAP solutions have a zero eigenvalue at any value of the free energy. When the free energy goes to the equilibrium value the continuous band develops itself a zero eigenvalue and it is interesting to ask if the two eigenvectors, the one of the continuous part and the one of the isolated eigenvalue, are different.  

Let us reconsider the expression for the determinant of the matrix $A$ defined above in equation (\ref{abp}). We rewrite it in a way that takes into account the possibility that $\det B=0$.
\begin{equation}
\det A= \det B +\langle m | C | m \rangle \ ,
\end{equation}
where $C$ is the matrix of the cofactors of $B$ such that $B^{-1}=C/\det B$.
If $\det B$ is different from zero the condition for the vanishing of $\det A$ is
that the l.h.s. of equation (\ref{ABinv}) vanishes and the eigenvalue of $A$ with zero eigenvector is $B^{-1}|m\rangle$.
Instead  if $\det B=0$  the condition for the vanishing of $\det A$ is simply
\begin{equation}
\langle m | e_0 \rangle=0  \ ,
\label{me0}
\end{equation}
where $| e_0 \rangle$ is the eigenvector of $B$ with zero eigenvalue; the previous result can be  simply derived considering the form of the matrix $C$ in a basis in which $B$ is diagonal.  
The last equation implies that the eigenvector $|e_0\rangle$ is also an eigenvector of $A$ with zero eigenvalue. Instead the vector $C|m\rangle$ is simply zero.
To understand what is the behavior of the eigenvector of the isolated eigenvalue it is convenient to study the situation in which the matrix $B$ has a small eigenvalue equal to $\epsilon$ corresponding to an eigenvector $|e_\epsilon\rangle$. In this case one finds that the eigenvector of the matrix $A$ corresponding to the vanishing isolated eigenvector is equal to
\begin{equation}
B^{-1} | m \rangle \propto |e_\epsilon\rangle+ O(\sqrt{\epsilon}) \ .
\end{equation}
As an eigenvalue of $B$ approaches zero the eigenvector $B^{-1} | m \rangle $ of $A$ becomes equal to the eigenvector of $B$ with zero eigenvalue.
The previous arguments can be extended in the case when there are more than one zero eigenvalue in the matrix $B$. If the matrix $B$ has $k$ eigenvectors $|e_i\rangle$ with zero eigenvalues  the matrix $A$ has at least $k-1$ eigenvectors with zero eigenvalue. Instead if the condition  (\ref{me0}) holds for all the $k$ eigenvectors the matrix $A$ has $k$ eigenvectors with zero eigenvalue.
Finally let us note that the condition (\ref{me0}) implies that the eigenvector with zero eigenvalues of the TAP solutions corresponding to the equilibrium states is orthonormal to the vector of the magnetizations.

\section{Conclusions}

In the introduction we saw that by dropping the modulus of the determinant we face two problems.
Assuming that the BM theory gives the actual complexity (the one {\it with} the modulus)
the problem is to obtain the exact result for the sum of solutions weighted with the sign of the determinant; for instance the sum over all solutions must yield $1$  because of the Morse theorem. In \cite{K91} Kurchan suggested that this is recovered in the BM theory  due to the vanishing of the prefactor at all order in $1/N$. Indeed if the prefactor is proportional to $e^{-N \Sigma_{BM}}$ (such that $e^{-N \Sigma_{BM}}e^{N \Sigma_{BM}}=1$) the resulting expansion in powers of $1/N$ is zero at all orders. According to our results the prefactor vanishes at all orders  not only for the total number of solutions but separately at any given free energy $f$, therefore, applying the same argument, we must conclude that at any given free energy there is an equal number of solutions with positive and negative determinant, much as if the Morse theorem be valid at any $f$ separately.

The previous arguments imply that the BM saddle point describes solutions with both positive and negative Hessian at all free energies. On the other hand the leading part of the spectrum of the TAP Hessian is strictly positive in the BM theory and the question is: where are the negative Hessian solutions? Recently Aspelmeier, Bray and Moore \cite{ABM} have shown that a solution to this problem can be found noticing that the TAP spectrum contains a single isolated eigenvalue besides the continuous strictly positive band. They have checked numerically that this eigenvalue is zero in the BM theory. Therefore in the thermodynamic limit the Morse theorem is recovered because the solutions have zero determinant. At finite $N$ the isolated eigenvalue acquires a  non-zero value $O(1/\sqrt{N})$, each solutions becomes a minimum or a saddle and at the same time another solution with an opposite value of the isolated eigenvalue develops at a distance $O(1/\sqrt{N})$. These results are valid for the total number of solutions (consistently with the Morse theorem) but also with the solutions at any given free energy consistently with the conclusions of the previous paragraph.
Here we have proved that the vanishing of the isolated eigenvalue at any given free energy is an exact result that is a consequence of SUSY breaking.

Since it has been shown \cite{noiquen} that the SUSY solution is inconsistent (except at the lower-band edge, see \cite{BMY,noiquen}) in full replica-symmetry-breaking (FRSB) models like the SK  model (at variance with 1RSB models), the BM solution remains the only viable candidate to describe a finite complexity in these models and seems to be supported by recent numerical results \cite{CGP04}. The SUSY approach has given us further insight into this non-SUSY theory, due to the presence of the isolated eigenvalue founded in \cite{ABM} we know that the TAP solutions described by the BM saddle point are inflection points in the TAP landscape while they split in couples minima-saddles at finite $N$. These features may be of some importance for the dynamics. 
Note that the presence of zero eigenvalues in the spectrum is a well-known feature of the lowest TAP solutions but has a different origin, indeed it is due to the fact that at the lower band edge the continuous band of eigenvalues extends down to zero. We have considered the problem of the relation between the eigenvector associated to the vanishing isolated eigenvalue and the eigenvector associated to the zero eigenvalue coming from the continuous band . In the end of section (\ref{sec:theiseigen}) we have shown that the two eigenvectors concide.
However in this case the SUSY approach yields  the non-trivial result that the eigenvector of the Hessian corresponding to the zero eigenvalue is orthonormal to the vector ot the magnetization $\{m\}$, according to equation (\ref{me0}).
  
An open problem of the BM theory is how to compute the $1/N$ correction to the number of solutions. A possible way is to remove the effect of the vanishing isolated eigenvaue responsible for the cancellation of the prefactor. Work is in progress in this direction and on the extension of the results presented here to the replicated case. 
\\
{\bf Aknowledgements}. We wish to thank A. Cavagna, A. Crisanti, I. Giardina and L. Leuzzi for useful discussions. We also thank A. Bray and M. Moore for some stimulating comments on the last part of the paper.

\end{document}